\documentclass[%
reprint,
superscriptaddress,
showpacs,preprintnumbers,
showkeys,
 amsmath,amssymb,
 aps,
]{revtex4-1}

\usepackage{graphicx}
\usepackage{dcolumn}
\usepackage{bm}

\def\be{\begin{equation}}
\def\ee{\end{equation}}
\def\bea{\begin{eqnarray}}
\def\eea{\end{eqnarray}}
\newcommand{\ket}[1]{\mbox{$|#1\rangle$}}

\begin{document}

\title{A Robust Quantum Random Access Memory}

\author{Fang-Yu Hong}
\affiliation{Department of Physics, Center for Optoelectronics Materials and Devices, Zhejiang Sci-Tech University,  Hangzhou, Zhejiang 310018, China}
\author{Yang Xiang}
\affiliation{School of Physics and Electronics, Henan University, Kaifeng, Henan 475004, China}
\author{Zhi-Yan Zhu}
\affiliation{Department of Physics, Center for Optoelectronics Materials and Devices, Zhejiang Sci-Tech University,  Hangzhou, Zhejiang 310018, China}
\author{Li-zhen Jiang}
\affiliation{College of Information and  Electronic Engineering, Zhejiang Gongshang University, Hangzhou, Zhejiang 310018,China}
\author{Liang-neng Wu}
\affiliation{College of Science, China Jiliang University, Hangzhou, Zhejiang 310018, China}
\date{\today}
\begin{abstract}
 A ``bucket brigade" architecture for a quantum random memory of $N=2^n$ memory cells needs $n(n+5)/2$ times of quantum manipulation on control circuit nodes per memory call. Here we propose a scheme, in which only average $n/2$ times manipulation  is required to accomplish a memory call. This scheme may significantly decrease  the time spent on a memory  call  and  the average overall error rate per memory call. A physical implementation scheme for storing an arbitrary state in a selected memory cell followed by reading it out is discussed.
\end{abstract}


\maketitle

A random access memory (RAM) is a fundamental computing device, in which information (bit) can be stored in any memory cell and be read out at discretion \cite{rfey,rjtb}. A RAM is made up of an input address register, a data register, an array of memory cells, and a controlling circuit.  A unique address is ascribed to each memory cell. When the address of a memory cell is loaded into the address register, the memory cell is selected and the information in the data register can be stored in it or the information of the cell can be read out to the data register. Like its classic counterpart, quantum random access memory (QRAM) is the building block of large quantum computers. A QRAM is a RAM  working in a way with quantum characteristic: the address and  data registers are comprised of qubits instead of bits, and every node of the controlling circuit is composed of a quantum object.  When the address state is in superposition, $\sum_i\alpha_i\ket{x_i}$, the read-out operation gives the output state $\sum_i\alpha_i\ket{q_i}$ in the data register, where $\ket{q_i}$ is the quantum information stored in the memory cell $i$ associated with the address $\ket{x_i}$. Quantum random access memories storing classic data can exponentially speed up the pattern recognition \cite{gsrs,rsch,catr,dcdm}, discrete logarithm \cite{aamb,amch}, and quantum Fourier transform, and quantum searching on a classical database \cite{mnic}. A general QRAM is an indispensable for the performance of many algorithms, such as quantum searching \cite{lkgr}, element distinctness \cite{aamba,amchi}, collision finding \cite{gbph}, general NAND-tree evaluation \cite{amchil}, and signal routing \cite{vgsl}.

 In a seminal paper \cite{vgsl,vglm}, Giovannetti {\it et al.} (GLM) proposed a promising bucket-brigade architecture for  QRAMs, which exponentially reduce the requirements for a memory call.  However, in GLM scheme, $n$ times of quantum unitary transformations per memory call is required to turn one quantum trit initialized in $\ket{wait}$ in each node of the control circuit into  $\ket{left}$ or $\ket{right}$,  and all flying qubits  including  address qubit and bus qubit can pass through an arbitrary node of the controlling circuit only if  a successful quantum manipulation has been performed on the trits, leading to the times of manipulations on the nodes $N_c=n(n+5)/2$ per memory call for $2^n$ memory cells, where $n$ is the number of bits in the address register. Here we present a QRAM scheme, where the quantum object in every node have only two possible states $\ket{left}$ and $\ket{right}$. On average the times of quantum manipulations on the nodes per memory call can be reduced to $N_c=n/2$, significantly decreasing  both the decoherence rate and the time spent on a QRAM call. A physical implementation  for information storage and read-out on a QRAM is presented.

\begin{figure}[t]
\includegraphics[width=8cm]{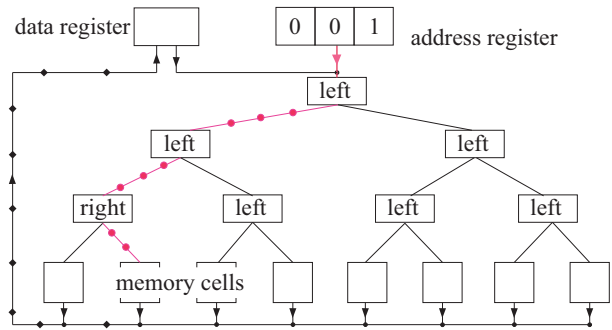}
\caption{\label{fig1}(color online). Schematics for a quantum random access memory. In each node of the binary  control circuit, a qubit in $\ket{right}(\ket{left})$ routes the approaching signals right (left). A single photon can excite the qubit from  $\ket{left}$ to $\ket{right}$ with the aid of a classical impedance-matched control field $\Omega(t)$. Here the third level memory cell $\ket{001}$ is addressed through the selected path marked with red circles. The read-out state is transferred to a data register along the path marked with black squares.}
\end{figure}

 The main idea is shown in Fig.\ref{fig1}. The $N$ memory cells are positioned at the end of a bifurcation control circuit with $n=\log_2 N$ levels. At each node of the control circuit there is a qubit with two states  $\ket{left}$ and $\ket{ right}$. The state of the $j$th qubit in the address register controls which route to follow when a signal arrives at a node in the $j$th level of the circuit: if the node qubit is $\ket{0}$, the left path is chosen; if it is  $\ket{1}$, the right path is chosen. For example, an address register $\ket{001}$ means that left at the 0th level, left at the next, and right at the second.  Illuminated by a control laser pulse  a node qubit in state  $\ket{left}$ will flip to $\ket{right}$  if the incoming address qubit is $\ket{1}$, or remain in $\ket{left}$ if the address qubit is $\ket{0}$. Without the control pulse, a node qubit in $\ket{left}$ ($\ket{right}$)will deviate any incoming signal along the left(right) side route.

 First, all the node qubits are initialized in state $\ket{left}$. Then the first qubit of the address register is dispatched through the circuit. At the first node, the address qubit  incurs a unitary transformation $U$ on  the node qubit with the help of  a control pulse $\Omega(t)$: $U\ket{0}\ket{left}=\ket{0}\ket{left}$ and $U\ket{1}\ket{left}=\ket{0}\ket{right}$. Next the second qubit of the address register is dispatched through the circuit, follow left or right route relying on the state of the first node qubit, and arrives at one of the two nodes on the second level of the circuit. The node qubit illuminated by the control pulse  will make a corresponding state change according to the state of the second address qubit, and so on. Note that the $i$th control pulse $\Omega(t)$ address all the nodes of the $i$th level control circuit simultaneously.   After all the $n$ qubits of the address register have gone through the whole circuit, a unique path of $n$ qubits has been singled out from the circuit (see Fig.\ref{fig1}). Subsequently, a single photon is sent along the selected path to single out a memory cell. After that an arbitrary unknown state in the data register can be transferred to the selected memory cell along the selected path, or  the state of the selected memory cell can be read out to the data register along the path with black squares in  Fig.\ref{fig1}. Finally, all the node qubits are reset to $\ket{left}$ for a next memory address.

Because the state of a node qubit $\ket{left}$ will not be affected by the  control pulse illuminating  should the qubit of the address register be in state $\ket{0}$, on average there are $n/2$ node qubits will flip to $\ket{right}$ in each memory call. This means that on average only $n/2$ times of control manipulations are really performed in each memory call. As a result, the mean comprehensive error rate per memory address is $n\epsilon/2=\frac{1}{2}\text{log}_2N \epsilon$  with the assumed error rate $\epsilon$ per node qubit flip event. In contrast, the GLM scheme requires $n$ times state flip for a memory call.  In addition, in the GLM scheme a photon may pass through a node only when a control pulse is applied on the quantum trit, resulting the overall times of  quantum manipulations on the quantum trits per memory call be $n(n+5)2$, which has included $2n$ times of manipulations  for a signal photon going  to a memory cell and  back to a data register along a same selected path. Here  a single-photon can pass through a node without any quantum manipulation, therefore the average times of  quantum manipulation really performed on  node qubits per memory call is $n/2$. Thus this scheme may significantly decrease the average overall error rate  and shorten the time required for a memory call.

\begin{figure}[t]
\includegraphics[width=8cm]{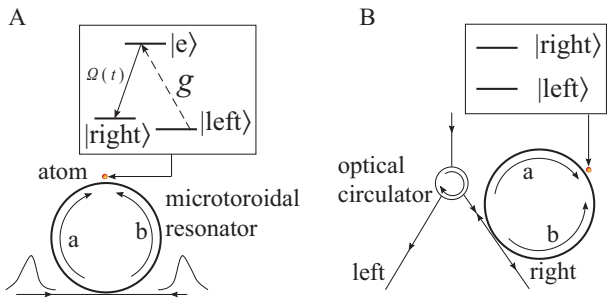}
\caption{\label{fig2} (Color online). Schematic diagram of a node consisting of a three-level atom and a microtoroidal resonator. (A) An address qubit consisting of zero or one photon is split equally in counter-propagating directions and coherently stored using an impedance-matched control field $\Omega(t)$, leading to a state flip of the atom conditioned on the photon number. (B) By employing an optical circulator an photon travels along a tapered fiber being in critical coupling  to the resonator will go along the left (right) path if the atom is in state $\ket{left}(\ket{right})$.}
\end{figure}

Now we discuss a physical implementation. The node qubit is encode on an atom with level $\ket{left}$, $\ket{right}$, and an intermediate state $\ket{e}$ (see Fig.\ref{fig2} A). The transition between $\ket{left}$ and $\ket{e}$ is coupled to the evanescent fields of modes $a$ and $b$ of of frequency $\omega_c$ of a microtoroidal resonator. State $\ket{right}$ is coupled to $\ket{e}$ by classical control field $\Omega(t)$. A tapered fiber and the resonator are assumed to be in critical coupling  where the input photons of frequency $\omega_p=\omega_c$ are all reflected back and the forward flux in the fiber drops to zero when the atom transition ($\ket{left}\rightarrow\ket{e}$) is far detuned from the resonator frequency $\omega_c$ \cite{tabd}. If the atomic transition ($\ket{right}\rightarrow\ket{e}$) is on resonant with the resonator, the input photons can transmit the resonator and travel forward one by one \cite{bdap}.   A single-photon can be coherently stored in the atom initialized in $\ket{left}$ by applying the control pulse $\Omega(t)$ simultaneous with the arrive of the photon which is equally divided and incident from both sides of the tapered fiber simultaneously (see Fig.\ref{fig2}) \cite{hong}.  The photon storage results in a state flip of the atom to $\ket{right}$. If no single-photon is contained in the incoming field, the atom does not affected by the control pulse  illuminating and remains in $\ket{left}$ \cite{hong}.

The switch function of a node qubit in a QRAM can be realized as follows:  first, the address qubit is encoded as $\alpha\ket{0}_p+\beta\ket{1}_p$ with Fock state $\ket{n}_p(n=0,1)$ and arbitrary unknown complex coefficients $\alpha$ and $\beta$; the first qubit of the address register is sent out along the control circuit and is coherently stored in the atom  in the first node by applying the control pulse $\Omega(t)$ simultaneous with the arrival of the address qubit  equally split and incident from both sides of the tapered fiber simultaneously. This storing process will incur a state flip of the node atom to $\ket{right}$  if the address qubit is $\ket{1}$, or make no change in the atom state $\ket{left}$ if the address qubit is $\ket{0}$. When the second address qubit is sent out and meet the first node, it will be reflected back and travel along the left path by applying an optical circulator in one side of the tapered fiber (see Fig.\ref{fig2} b) if the atom in the first node is in $\ket{left}$, or will transmit the resonator and go along the right path if the atom is in $\ket{right}$. When the second address qubit arrive at one of the two nodes on the second level, it will be coherently stored in the node atom  and left the atom in $\ket{left}$ or $\ket{right}$ dependent on the photon number contained in the address qubit, and so on.

\begin{figure}[t]
\includegraphics[width=8cm]{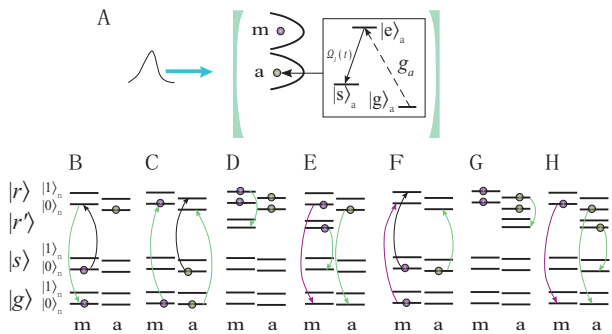}
\caption{\label{fig3}(color online). Schematics for writing to and reading out of  a memory cell. (A) A single-photon can be coherently stored in a memory cell consisting of two atoms $m$ and $a$ positioned inside of a high-Q cavity with a time-dependent control pulse $\Omega_2(t)$. Diagrams of energy levels and pulses sequences for storing unknown state to the selected memory cell (B to E) and for reading out of the content of the selected memory cell (F to H). }
\end{figure}

We assume that each quantum memory cell in the memory array  consists of a memory atom $m$ and an ancillary atom $a$, which are confined in two harmonic traps and  positioned  inside a high quality cavity (see Fig.\ref{fig3}A). The ancillary atom  has  a three-level structure: $\ket{g}_a$ is coupled to $\ket{e}_a$ by the   field of the cavity mode with strength $g_a$; $\ket{s}_a$ is coupled to $\ket{e}_a$ by a classic control field $\Omega_1(t)$, where  the subscript $a$ denotes an ancillary atom.   After a path  to the memory array is singled out, a single-photon is sent along the path  to the selected memory cell. A control laser pulse  $\Omega_1(t)$ is applied to the ancillary atoms initialized in state $\ket{g}_a$ at the moment when the photon arrives at the memory cell, resulting a state flip of  atom a to $\ket{s}_a$ \cite{wyrl,wyrll}. To avoid to be involved into the quantum operations aimed on the selected memory cell, the ancillary atoms non-selected are excited to a stable state $\ket{t}_a$ by a $\pi$ pulses on transition $\ket{g}_a\rightarrow\ket{t}_a$.  Next we can do some quantum manipulations either to save an arbitrary unknown quantum state in the selected memory cell or to read out its content.

In the first place, we discuss how to save an arbitrary unknown quantum state in the memory cell identified by the address register. First, an initialization operation on the memory cell array is performed. This can be realized as follows:   atom a in state $\ket{s}_a$ is excited to a Rydberg state $\ket{r}_a$ by two-photon stimulated Raman (TWSR) pulses \cite{eutj}; TWSR pulses on the transition $\ket{s}_m\rightarrow \ket{r}_m$ (in terms of the perturbed state ) of the memory atoms are applied, resulting the selected memory atom  being excited to $\ket{r}_m$ and immediately flipping to the ground state $\ket{g}_m$ through spontaneous radiation (see Fig.\ref{fig3}B).  In this way only the memory atom in the selected memory cell is reset in  state $\ket{g}_m$, leaving the content of the others unchanged, because the states $\ket{r}_m $  of the non-selected memory atoms are off resonant with the TWSR pulses due to the absence of the strong Rydberg dipole interactions \cite{djjc}. The ancillary atom in Rydgerg $\ket{r}_a$ is brought to the ground state $\ket{g}_a$ by applying a $\pi$ pulse on its transition $\ket{r}_a\rightarrow\ket{g}_a$.

 Second, an arbitrary unknown state $\alpha\ket{0}_p+\beta\ket{1}_p$ with Fock bases $\ket{n}_p(n=0,1)$ is transferred along the selected path to the memory cells. On the arrival of the signal a classic control pulse $\Omega_2(t)$ is applied on the ancillary atoms initialized in the ground state $\ket{g}_a$, leading to a state map $ (\alpha\ket{0}_p+\beta\ket{1}_p)\ket{g}_a\rightarrow\ket{0}_p(\alpha\ket{g}_a+\beta\ket{s}_a)$ \cite{wyrl,wyrll}.

 State $ \alpha\ket{g}_a+\beta\ket{s}_a$ is then transferred to a memory atom by employing the strong dipole interaction between two Rydberg atoms \cite{hongx,pstr}. When both of the memory atom and the ancillary atom are in Rydberg states, the strong dipole interaction between them will couple their motion, which is best described in the basis of normal modes $\ket{j}_n(j=0,1,2,...)$. All the motion modes of the atoms are initially cooled to near their ground state by Raman sideband cooling on them \cite{cmdm}.  
 
 Third, we drive a $\pi$ pulse on transition on $\ket{g}_m\rightarrow\ket{r}_m$ on memory atoms to excite the memory  atoms from state $\ket{g}_m$ to state $\ket{r}_m$. Fourth, a $\pi$ pulse on transition $\ket{g}_a\rightarrow\ket{r}_a\ket{0}_n$ and a blue sideband (BSB) pulse on transition $\ket{s}_a\rightarrow\ket{r}_a\ket{1}_n$  are applied on the ancillary atoms, leading to  state
\be\label{eq1}
\ket{\psi}_1=\alpha\ket{r}_a\ket{r}_m\ket{0}_n+\beta\ket{r}_a\ket{r}_m\ket{1}_n
\ee
for the selected memory cell (see Fig. \ref{fig3}bB) and  state
$ \ket{\psi}_2=(\alpha_x\ket{r}_m+\beta_x\ket{s}_m)\ket{t}_a $ for other memory cells with their initial states state $\alpha_x\ket{g}_m+\beta_x\ket{s}_m$ (see Fig.\ref{fig3}C).  Here we have used the fact that the normal mode of motion is shared by the memory atom and the ancillary atom, both of which are in Rydberg states. Fifth, a $\pi$ pulse on the transition $\ket{r}_m\rightarrow \ket{g}_m$  unperturbed by the strong dipole interaction between two Rydberg atoms is applied on the non-selected memory atoms to restore them to their initial states $\alpha_x\ket{r}_m+\beta_x\ket{s}_m$.

 Sixth, a $\pi$ pulse on transition $\ket{r}_m\ket{1}_n\rightarrow\ket{r'}_m\ket{0}_n$ illuminates the memory atoms, resulting a state mapping
\be\label{eq3}
\ket{\psi}_1\rightarrow\ket{\psi}_2=\alpha\ket{r}_a\ket{r}_m\ket{0}_n+\beta\ket{r}_a\ket{r'}_m\ket{0}_n.
\ee
 Seventh, two $\pi$ pulses on  $\ket{r'}_m\ket{0}_n\rightarrow\ket{s}$ and $\ket{r}_m\ket{0}_n\rightarrow\ket{g}_m$, respectively, are applied on the memory atom, leading to an unitary transformation $
\ket{\psi}_2\rightarrow\ket{\psi}_3=(\alpha\ket{g}_m+\beta\ket{s}_m)\ket{r}_a$;
   the unknown state $\alpha\ket{0}+\beta\ket{1}$ has been stored on the selected memory atom. Note that this two pulses will not affect the states of  the non-selected memory atoms since the pulses is detuned from the transitions $\ket{r'}_m\rightarrow\ket{s}_m$ and $\ket{r}_m\rightarrow\ket{g}_m$, which are free of the influence of strong dipole interaction between two Rydberg atoms. Eighth, the ancillary atoms is restored  to the ground state $\ket{g}_a$ by a a $\pi$ pulse on the transition $\ket{r}_a\rightarrow\ket{g}_a$ (see Fig.\ref{fig3}E), and the non-selected ancillary atoms are restored to the ground state $\ket{g}$ for the next memory call by flipping to a intermediate level with a pulse and then falling to $\ket{g}$ through spontaneous radiation. In this way an arbitrary  unknown state can be transferred to the selected memory atom, leaving the states of other memory atoms unchanged.

 Now we discuss how to read out the content of the selected memory atom $\alpha_x\ket{g}_m+\beta_x\ket{s}_m$. First, a $\pi$ pulse on transition $\ket{s}_a\rightarrow\ket{r}_a$ is employed to excite the selected ancillary atom to state $\ket{r}_a$. Second, we employ two $\pi$ pulses on transition $\ket{g}_m\rightarrow\ket{r}_m\ket{0}_n$ and $\ket{s}_m\rightarrow\ket{r}_m\ket{1}_n$, respectively (see Fig.\ref{fig3}F), driving the  system of the  selected atoms $m$ and $a $ into  state
 \be\label{eq4}
 \ket{\psi}_4=\alpha_x\ket{r}_m\ket{r}_a\ket{0}_n+\beta_x\ket{r}_m\ket{r}_a\ket{1}_n.
 \ee

Third, a $\pi$ pulse on transition $\ket{r}_a\ket{1}_n\rightarrow\ket{r'}_a\ket{0}_n$ drive the selected system into  state
$\ket{r}_m(\alpha_x\ket{r}_a+\beta_x\ket{r'}_a)\ket{0}_n $ (see Fig.\ref{fig3}G).
 Fourth, the selected memory atom is sent to the ground state by a $\pi$ pulse on the transition $\ket{r}_m\ket{0}_n\rightarrow\ket{g}_m$. Fifth, two pulses on the transition $\ket{r}_a\rightarrow\ket{g}_a$ and $\ket{r'}_a\rightarrow \ket{s}_a$, respectively, set the selected ancillary atom in state $\alpha_x\ket{g}_a+\beta_x\ket{s}_a$ (see Fig.\ref{fig3}H); the content of the select memory atom is transferred to the ancillary atom. Note that  these  pulses do not influence  the non-selected atoms $a$ and $m$  because they have no strong dipole interaction. By applying a classical impedance-matched control field $\Omega(t)$, a matter-photon mapping $(\alpha_x\ket{g}_a+\beta_x\ket{s}_a)\ket{0}_p\rightarrow\ket{g}_a(\alpha_x\ket{0}_p+\beta_x\ket{1}_p)$
can be accomplished \cite{wyrl,wyrll}, transferring the state of the selected memory cell  to the flying qubit and leaving the ancillary atom in state $\ket{g}_a$. The flying qubit goes along the path with black squares to the data register (see Fig.\ref{fig1}). Finally, the non-selected ancillary atoms are initialized to state $\ket{g}_a$ for a next task.

In summary, we have presented a scheme  for  a quantum random access memory. With three-level memory system been substituted by a qubit in every node of the control circuit, this structure may significantly reduce overall  error rate per memory address and the memory address time. In addition, we have discussed a physical implementation based on microtoroidal resonator and strong-dipole interaction between two Rydberg atoms for a QRAM writing and read-out. The microtoroidal resonator and the tapered fiber may be replaced by a surface plasmon propagating on the surface of a nanowire-conductor-dielectric interface \cite{dcas,hongs}.


 This work was supported by the National Natural Science Foundation of China ( 11072218, and 11005031), by Zhejiang Provincial Natural Science Foundation of China (Grant No. Y6110314 and Y6100421), and  by Scientific Research Fund of Zhejiang Provincial Education Department (Grant No. Y200909693 and Y200906669).

\end{document}